\documentclass[%
reprint,
amsmath,amssymb,
aps,
prapplied,
]{revtex4-2}

\usepackage{graphicx}% Include figure files
\usepackage{dcolumn}% Align table columns on decimal point
\usepackage{bm}% bold math
\usepackage[colorlinks,linkcolor=blue,anchorcolor=blue,citecolor=blue]{hyperref}

\begin{document}

\preprint{APS/123-QED}

\title{Trusted source noise model of discrete-modulated continuous-variable quantum key distribution
}% Force line breaks with \\

\author{Mingze Wu$^{1}$}
% \altaffiliation[Also at ]{Physics Department, XYZ University.}%Lines break automatically or can be forced with \\
% \thanks{Correspondence: zhangyc@bupt.edu.cn.}%
\author{Junhui Li$^{1}$}
\email{Correspondence: lijunhuiphy@bupt.edu.cn}
\author{Bingjie Xu$^{2}$}
\author{Song Yu$^{1}$}
\author{Yichen Zhang$^{1,}$}%
\email{Correspondence: zhangyc@bupt.edu.cn}

\affiliation {$^{1}$State Key Laboratory of Information Photonics and Optical Communications, School of Electronic Engineering, Beijing University of Posts and Telecommunications, Beijing 100876, China\\
$^{2}$Science and Technology on Security Communication Laboratory, Institute of Southwestern Communication, Chengdu 610041, China}

%\collaboration{MUSO Collaboration}%\noaffiliation

%\author{Charlie Author}
 %\homepage{http://www.Second.institution.edu/~Charlie.Author}
%\affiliation{ Second institution and/or address\\ This line break forced% with \\}%
%\affiliation{Third institution, the second for Charlie Author}%
%\author{Delta Author}
%\affiliation{%Authors' institution and/or address\\This line break forced with \textbackslash\textbackslash}%

%\collaboration{CLEO Collaboration}%\noaffiliation

\date{\today}% It is always \today, today,
             %  but any date may be explicitly specified

\begin{abstract}

Discrete-modulated continuous-variable quantum key distribution offers a pragmatic solution, greatly simplifying experimental procedures while retaining robust integration with classical optical communication. Theoretical analyses have progressively validated the comprehensive security of this protocol, paving the way for practical experimentation. However, imperfect source in practical implementations introduce noise. The traditional approach is to assume that eavesdroppers can control all of the source noise, which overestimates the ability of eavesdroppers and underestimates secret key rate. In fact, some parts of source noise are intrinsic and cannot be manipulated by eavesdropper, so they can be seen as trusted noise. We tailor a trusted model specifically for the discrete-modulated protocol and upgrade the security analysis accordingly. Simulation results demonstrate that this approach successfully mitigates negative impact of imperfect source on system performance while maintaining security of the protocol. Furthermore, our method can be used in conjunction with trusted detector noise model, effectively reducing the influence of both source and detector noise in experimental setup. This is a meaningful contribution to the practical deployment of discrete-modulated continuous-variable quantum key distribution systems.

%\begin{description}
%\item[Usage]
%Secondary publications and information retrieval purposes.
%\item[Structure]
%You may use the \texttt{description} environment to structure your abstract;
%use the optional argument of the \verb+\item+ command to give the category of each item.
%\end{description}
\end{abstract}

%\keywords{Suggested keywords}%Use showkeys class option if keyword
                              %display desired

\maketitle

%\keywords{Suggested keywords}%Use showkeys class option if keyword
                              %display desired

%\tableofcontents

\section{\label{set1}Introduction}
Quantum key distribution (QKD)\ \cite{1984Quantum} is one of the significant applications in the field of quantum information, possessing strong practicability and commercial value\ \cite{pirandola2020advances,xu2020secure,portmann2022security}. It enables two trusted users to establish secret keys with unconditional security in theory. Continuous-variable quantum key distribution (CV-QKD) using coherent states\ \cite{ralph1999continuous,grosshans2002continuous,weedbrook2004quantum} exhibits strong compatibility with classical optical communications, and can achieve high secret key rates over short to medium distances, thus possesses significant advantages in practical deployments within metropolitan areas\ \cite{zhang2024continuous}. In recent years, theory\ \cite{leverrier2015composable,leverrier2017security,pirandola2021composable} and experimentation\ \cite{zhang2020long,tian2022experimental,hajomer2024long} of CV-QKD have been developed. Advancements in chip integration\ \cite{zhang2019integrated,li2023continuous,hajomer2023continuous,bian2024highly,bian2024Continuous}, and network\ \cite{du2023continuous,bian2023high,hajomer2024continuous,pan2024high} have significantly enhanced the practicality of CV-QKD.

Distinguished by modulation methods, CV-QKD can be categorized into Gaussian-modulated protocols\ \cite{ralph1999continuous,grosshans2002continuous,weedbrook2004quantum}, discrete-modulated protocols\ \cite{leverrier2009,li2018user,leverrier2019,lin2019,matsuura2021finite}, etc. Gaussian-modulated protocols are proposed earlier and have achieved considerable development due to the simplicity of security proofs taking advantage of the extremality of Gaussian states\ \cite{wolf2006extremality,garcia2006unconditional,navascues2006optimality}. However, Gaussian modulation requires thousands of discrete constellations to approximate continuous modulation in practical experiments\ \cite{jouguet2012}, which will consume more random numbers and post-processing resource. In contrast, discrete modulation, a commonly used modulation method in classical optical communication, is straightforward to implement and highly practical, representing one of the promising directions for CV-QKD\ \cite{leverrier2009}. Due to the non Gaussian property of the quantum states shared by both communication parties under discrete modulation, security proof of discrete-modulated protocol is limited by the assumption of linear channel for a long time\ \cite{leverrier2009}. Fortunately, introduction of Semidefinite programming (SDP) method has successfully resolved this issue\ \cite{leverrier2019,lin2019}. The authors in\ \cite{leverrier2019} realize asymptotic security analysis of quadrature-phase-shift-keying (QPSK) modulation protocol, and use SDP to constrain the covariance term in the covariance matrix. After that, the method is further extended to arbitrary modulation, enabling analytical bound of SDP\ \cite{denys2021explicit}. Additionally, experimental verification has been successfully implemented\ \cite{wang2022sub,pan2022experimental,pereira2022probabilistic,roumestan2024shaped}. However, this method remains reliant on the optimality theorem of Gaussian state, resulting in the bound that is not tight enough. The authors in\ \cite{lin2019} use quantum relative entropy to express the security key rate, optimizing quantum relative entropy through SDP via a two-step numerical calculation method\ \cite{coles2016numerical,winick2018reliable} to obtain a more compact security key rate. This is a highly appealing method that has undergone further enhancements\ \cite{liu2021homodyne,upadhyaya2021dimension,kanitschar2022optimizing,upadhyaya2021dimension}, and has been thoroughly verified through long-distance experiments\ \cite{tian2023high}. Moreover, based on this method, finite-size security analysis of discrete-modulated protocol under collective attack\ \cite{kanitschar2023finite} and coherent attack\ \cite{bauml2023security} have also been proposed.

It is worth mentioning that most security analysis in previous works are solely based on ideal devices, which utilize ideal light sources, detectors, and other devices that conform to fundamental assumptions. Unfortunately, in practical systems, few devices can perfectly adhere to the basic assumptions of the protocols, thus necessitating adjustment of security analysis methods\ \cite{scarani2009security,laudenbach2018continuous}. Researches into non-idealities in Gaussian-modulated CV-QKD are quite extensive\ \cite{lodewyck2007quantum,fossier2009improvement,usenko2016trusted}. However, as we know, trust noise model on the source side of discrete-modulated protocols has not been studied. Following the approaches of Gaussian modulation, there have been some developments on non-ideal detectors of discrete-modulated protocols, including trusted noise model\ \cite{lin2020trusted} and quantum hacking attacks\ \cite{fan2023quantum}.

Besides detection and channel propagation, state preparation is also important in any QKD protocol. During this step, lasers, digital-to-analog converters, and modulators all exhibit various non-idealities. These non-idealities result in noise in the prepared quantum states. In previous security analysis, all the source noise is considered as untrusted excess noise which can be fully accessed by the eavesdropper, thus have huge impact on system performance\ \cite{jouguet2012analysis}. Nevertheless, some source noise originates from intrinsic stochastic mechanism, or always remains within Alice's system, actually cannot be manipulated by the eavesdroppers\ \cite{usenko2010,shen2011continuous}. Therefore, previous treatment to see all the source noise untrusted just exaggerates Eve’s power and leads to untight bound on security key rate\ \cite{shen2009security}. For Gaussian-modulated protocols, there are many researches on the imperfect state preparation, including trusted source noise model\ \cite{usenko2016trusted} and source monitoring\ \cite{yang2012source,chu2021}. Non-idealities also exist in discrete-modulated CV-QKD. However, due to the difference in security analysis methods, these source noise models cannot directly apply to discrete-modulated CV-QKD.

In this contribution, we provide the trusted source noise model of discrete-modulated CV-QKD protocol based on foregoing security analysis method with SDP.  We analyze the non-ideality in state preparation process of discrete-modulated protocol, and show that some parts of noise cannot be manipulated by eavesdroppers, so can be trusted. For trusted source noise, thermal state is used for modeling, and the trusted source noise model is established. According to this model, secret key rate of discrete-modulated CV-QKD is calculated, which shows that previous untrusted source noise treatment does significantly underestimate secret key rate. Numerical results also show that the secret key rate obtained by adjusting the security analysis based on trusted noise model can approach the key rate with ideal noiseless devices, indicating that trusted source noise model can almost eliminate the influence of trusted source noise on system performance.

Organization of this paper is as follow. In Sec.\ \ref{set2}, we delve into underlying causes and comprehensive influence of source noise during the state preparation process in a practical discrete-modulated CV-QKD system. In Sec.\ \ref{set3}, we describe discrete modulation CV-QKD protocol with noisy coherent states, and then give security analysis of the protocol in detail. In Sec.\ \ref{set4}, we present simulation results. Finally, we have some discussions and conclude the work that has been done in this paper in Sec.\ \ref{set5}.

\begin{figure}[t]
	\centering
	\includegraphics[width=8.5cm]{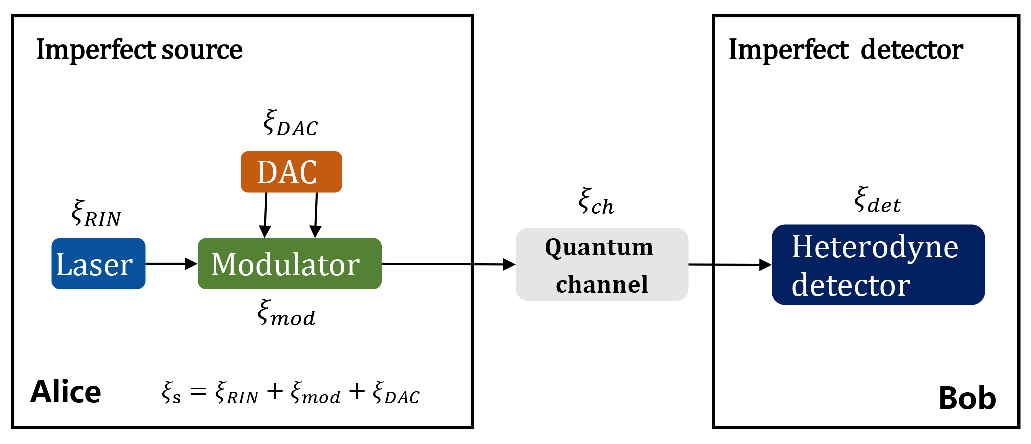}% Here is how to import EPS art
	\caption{\label{pm1}
		Practical discrete-modulated CV-QKD system with source noise introduced in Sec.\ \ref{set2}. $\xi_{\rm RIN}$: relative intensity noise, $\xi_{\rm mod}$: noise from modulator, $\xi_{\rm DAC}$: noise from DAC error, $\xi_{s}$: source noise, $\xi_{\rm ch}$: quantum channel noise, $\xi_{\rm det}$: detector noise.
	}
\end{figure}

\section{\label{set2}Imperfection in preparing quantum states}

\begin{figure}[t]
	\centering
	\includegraphics[width=7cm]{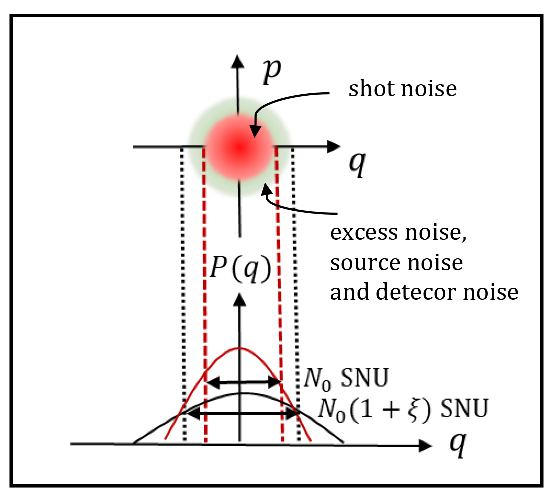}% Here is how to import EPS art
	\caption{\label{V}
		Phase-space representation of the coherent state received by Bob. The red circle represents shot noise with variance of $N_{0}$, the green outer ring represents excess noise, both source noise and detector noise, and the total variance of the noisy coherent state is $N_{0}(1+\xi)$.
	}
\end{figure}

In practical experiments, various imperfections introduce noise that can significantly compromise the system's performance. These unwanted noises, known as excess noise, can be quantified by extra fluctuation in the quadrature measurements. Assuming being decomposable, in practical discrete-modulated CV-QKD system, the excess noise can be attributed to various sources as shown in Fig.\ \ref{pm1}, including laser relative intensity noise $\xi_{\rm RIN}$, modulator noise $\xi_{\rm mod}$, digital-to-analog converter (DAC) noise $\xi_{\rm DAC}$, quantum channel noise $\xi_{\rm ch}$, detector noise $\xi_{\rm det}$, etc.\ \cite{laudenbach2018continuous},
\begin{equation}\label{noise}
   \xi = \xi_{\rm RIN}+\xi_{\rm mod}+\xi_{\rm DAC}+\xi_{\rm ch}+\xi_{\rm det}+...,
\end{equation}
where $\xi$ represents equivalent excess noise at source side. 
Since signal laser, modulator and DAC exist in the transmitter of CV-QKD system, this part of noise can be summed as source noise $\xi_{s}$
\begin{equation}\label{eq:0}
   \xi_{s} = \xi_{\rm RIN}+\xi_{\rm mod}+\xi_{\rm DAC}.
\end{equation}
The phase-space representation of the coherent state received by Bob with excess noise is shown in Fig.\ \ref{V}. The effect of source noise on coherent states is to expand the fluctuation range, that is, to increase the variance from $N_0$ to $N_0(1+\xi_s)$, where $N_0$ is shot noise variance. In the following, noise is represented in natural unit by default. The three types of source noise are described below respectively. The excess noise caused by relative intensity noise comes from the power fluctuations of laser
\begin{equation}\label{eq:1}
   \xi_{\rm RIN} = V_{M}\sqrt{\rm RIN \Delta\nu},
\end{equation}
where $V_{M}$ is the modulation variance of Alice's data, RIN is the relative intensity noise, and $\Delta\nu$ is the laser linewidth.
The excess noise of modulator mainly comes from modulated light extinction
\begin{equation}\label{eq:2}
	\xi_{\rm mod} = V_{M}10^{-d_{\rm dB}/10},
\end{equation}
where $d_{\rm dB}$ is the extinction ratio of the modulator. 
The excess noise introduced by the DAC arises from errors in converting signal digits into voltage, which can be bounded by
\begin{equation}\label{eq:3}
	\xi_{\rm DAC} \le V_{M}\left[\pi \frac{\delta U_{\rm DAC}}{U_{\rm DAC}}+\frac{1}{2}\left(\pi \frac{\delta U_{\rm DAC}}{U_{\rm DAC}}\right)^{2}\right]^{2},
\end{equation}
where $U_{\rm DAC}$ is the signal voltage, and $\delta U_{\rm DAC}$ is a specific deviation.

It is worth mentioning that not all of the excess noise in Eq. (\ref{noise}) can be used by eavesdroppers, such as source noise $\xi_s$ and detector noise $\xi_{\rm det}$, as they are within the system and no quantum hacking attack has been proposed to manipulate source noise, up to now. If they are treated as untrusted, they can lead to overestimation of the eavesdropper's capabilities, significantly reducing system performance. Therefore, if this part of the noise can to be trusted, system performance will improve. If the corresponding quantum hacking attack appears in the future, then the corresponding noise can be removed from the trusted part. The trusted noise model in our work can adapt to this change. Since trusted detector noise for discrete modulated CV-QKD has been studied in\ \cite{lin2020trusted}. Next, we mainly consider the trusted noise model of imperfect source and combine it with trusted detector noise.

\section{\label{set3}Discrete-modulated CV-QKD with imperfect source}
In this section,  we focus on practical discrete-modulated CV-QKD system with source noise in Fig.\ \ref{pm1}, taking QPSK modulated CV-QKD as an example. The protocol description is shown in Appendix \ref{A}, based on previous work\ \cite{lin2019,lin2020trusted}. We present the model of trusted source noise, together with comprehensive security analysis of the protocol. It is worth noting that our scheme modeling imperfect source can be scaled up to higher number of modulation constellations, enhancing protocol performance at the cost of increased computational complexity. For protocol using homodyne detection, our model is also applicable.

\begin{figure*}[t]
\centering
\includegraphics[width=15cm]{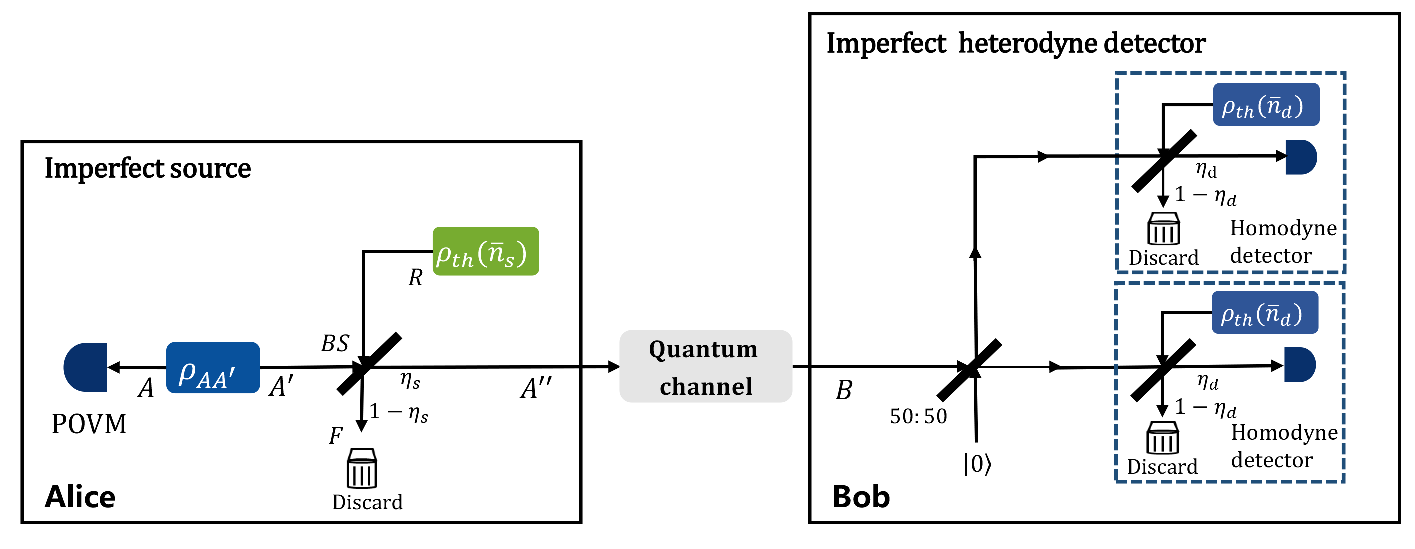}% Here is how to import EPS art
\caption{\label{1}
Schematic entanglement-based discrete-modulated CV-QKD protocol with imperfect source, where $\rho_{\rm th}(\bar{n}_s)$ is thermal state with average photon number $\bar{n}_s$, BS is beam splitter with transmittance $\eta_{s}=0.9999$, POVM is positive operator valued measurement $M^{A}$. Mode $F$ is discarded. Bob performs heterodyne detection with trusted detector noise\ \cite{lin2020trusted}.
}
\end{figure*}

%\subsubsection{Trusted source noise model}
Because the trusted part of source noise is inside Alice's system, it cannot be exploited by eavesdroppers and can be renamed as $\nu_{s}$ to distinguish from excess noise. In Gaussian-modulated protocol, source noise is modeled as a mode of an Einstein-Podolsky-Rosen (EPR) state, and the covariance matrix and optimality of Gaussian attacks are used in the security analysis\ \cite{usenko2010}. However, this approach is not applicable to discrete-modulated protocol. In our model, a beam splitter with transmittance $\eta_{s}\rightarrow1$ (we set $\eta_{s}=0.9999$) and a thermal state can be used to model the trusted part of source noise. The signal state $\rho_{AA'}$ and the thermal state $\rho_{\rm th}(\bar{n}_s)$ are coupled by a beam splitter, and the output state is transmitted to Bob through the quantum channel as shown in Fig.\ \ref{1}.  
When the source noise is $\nu_{s}$, the average photon number $\bar{n}_s$ of input equivalent thermal state $\rho_{\rm th}(\bar{n}_s)$ is $\bar{n}_s=\nu_s/(1-\eta_{s})N_{0}$, where $N_{0}$ is the variance of shot noise in unit of vacuum fluctuation. The variance of each quadrature of the thermal state is $\left[1+\nu_s/(1-\eta_{s})\right]N_{0}$.

In the ideal protocol, based on the source replacement scheme, when Alice sends coherent states $|\varphi_{k}\rangle$ with probability $p_k$ in the prepare-and-measure scheme, she equivalently prepares a bipartite state $|\Psi\rangle_{A A^{\prime}}$ in the entanglement-based scheme
\begin{equation}\label{eq:0000}
   |\Psi\rangle_{A A^{\prime}}=\sum_{x=0}^{3}{\sqrt{p_{x}}}|x\rangle_{A}|\varphi_{x}\rangle_{A^{\prime}},
\end{equation}
where $\left\{\left|x \right \rangle \right\}$ serves as an orthonormal basis set in register $A$. Density matrix of the bipartite state can be expressed as $\rho_{A A^{\prime}}=|\Psi\rangle_{A A^{\prime}}\langle\Psi|$. Alice reserves register $A$ and makes positive operator valued measurement (POVM) $M^{A}=\left\{M_x^A=\left|x\right\rangle
\left\langle x \right|:x\in\left\{0,1,2,3\right\}\right\}$ to obtain outcome $x$. Register $A'$ is coupled with the thermal state $\rho_{\rm th}(\bar{n}_s)$ through the beam splitter of transmittance $\eta_s$ to model the trusted source noise.
Density operator of thermal state with the average photon number $\bar{n}_s$ is
\begin{equation}\label{eq:0000}
\rho_{\mathrm{th}}(\bar{n}_s)=\sum_{n=0}^{\infty}\frac{\hat{n}^{n}}{(1+\hat{n})^{n+1}}|n\rangle\langle n|.
\end{equation}
In the coherent state representation, the thermal state can be rewritten as
\begin{equation}\label{eq:0000}
   \rho_{\mathrm{th}}(\overline{{{n}}}_s)=\frac{1}{\pi\bar{n}_s}\int_{\mathbb{C}}\exp(-\frac{|\beta|^{2}}{\bar{n}_s})\,|\beta\rangle\,\langle\beta|\,d^{2}\beta,
\end{equation}
where $\beta$ is the complex amplitude of a coherent state and $\mathbb{C}$ is the set of complex. After being coupled through the beam splitter, the quantum state of Alice's system, which includes the source noise, is
\begin{equation}\label{eq:0000}
\rho_{AA^{\prime\prime}}=\mathrm{Tr}_{R}[\hat{S}(\theta)\rho_{AA^{\prime}}\otimes\rho_{t h_{R}}(\overline{{{n}}}_s)\hat{S}
^{\dagger}(\theta)],
\end{equation}
where $\hat{S}(\theta)=\exp[\theta(\hat{a}_{A^{\prime}}^{\dagger}\hat{a}_{R}-\hat{a}_{A^{\prime}}\hat{a}_{R}^{\dagger})]$ is the beam splitter operator, $\hat{a}_{A^{\prime}}$(or $\hat{a}_{R}$) and $\hat{a}_{A^{\prime}}^{\dagger}$(or $\hat{a}_{R}^{\dagger}$) are respectively the annihilation and creation operator of system $A^{\prime}$(or $R$), and parameter $\theta$ is related to transmittance $\eta_{s}$ by $\eta_{s}=1/(1+\rm{tan}^{2}\theta)$. According to Appendix \ref{B}, if coherent states $|\alpha_{A'}\rangle$ and $|\alpha_{R}\rangle$ are coupled through a beam splitter of transmittance $\eta$, the output coherent state is 
\begin{equation}\label{eq:0000}
	\vert\alpha_{A''}\rangle=\vert\sqrt{\eta}\alpha_{A^{\prime}}+\sqrt{1-\eta}\alpha_{R}\rangle_{A''}.
\end{equation}

Thus, the quantum state of Alice's system
\begin{equation}\label{eq:0000}
\begin{aligned} 
&\rho_{A A^{\prime\prime}}=\frac{1}{\pi{\bar{n}_s}}\sum_{i,j=0}^{3}\sqrt{p_{i}p_{j}}|i\rangle_{A}\langle j|\int_{\mathbb{C}}\exp{\left(-\frac{|\beta|^{2}}{\bar{n}_s}\right)}\\
&|\sqrt{\eta}\alpha_{i}+\sqrt{1-\eta}\beta\rangle_{A^{\prime\prime}}\langle\sqrt{\eta_{A}}\alpha_{j}+\sqrt{1-\eta}\beta|d^{2}\beta,\\
\end{aligned}
\end{equation}
Then, register $A^{\prime\prime}$ is sent to Bob through the quantum channel. The quantum channel can be regarded as a completely positive and trace-preserving (CPTP) map ${\mathcal{E}}_{A^{\prime}\to B}$, which maps Alice's register $A^{\prime\prime}$ to Bob's register $B$, and is controlled by Eve. Therefore, the quantum state shared by Alice and Bob is
\begin{equation}\label{eq:0000}
\rho_{A B}=(\mathrm{id}_{A}\otimes{\mathcal{E}}_{A^{\prime}\to B})\rho_{A A^{\prime\prime}},
\end{equation}
where ${\mathrm{id}}_{A}$ refers to the identity channel within Alice's system $A$.

After Alice uses POVM $\left\{M_{x}^{A}\right\}$ to perform local measurement on register $A$ and gets the result $x$, the coherent state $|\varphi_{x}\rangle$ will be sent to Bob. Bob's received state conditioned on choice of $x$ is
\begin{equation}\label{eq:0000}
\rho_{B}^{x}=\frac{1}{p_{x}}\mathrm{Tr}_{A}[\rho_{A B}(|x\rangle\langle x|_{\it A}\otimes\mathrm{I}_{B})].
\end{equation}
Bob uses POVM $M^{B}=\left\{M_{y}^{B}\right\}$ to get his measurement results. For trusted detector noise, the POVM
of heterodyne detection $M_{y}^{B}=G_{y}$ can be expressed as\ \cite{lin2020trusted}
\begin{equation}\label{eq:0000}
G_{y}=\frac{1}{\eta_{d}\pi}\hat{D}\left(\frac{y}{\sqrt{\eta_{d}}}\right)\rho_{\mathrm{th}}\left(\frac{1-\eta_{d}+\nu_{\mathrm{el}}}{\eta_{d}}\right)\hat{D}^{\dagger}\left(\frac{y}{\sqrt{\eta_{d}}}\right),
\end{equation}
where $y\in\mathbb{C}$ is the complex amplitude of coherent state, $\eta_{d}$ is detector efficiency, $\nu_{\mathrm{el}}$ is detector electrical noise, $\hat{D}\left({y}/{\sqrt{\eta_{d}}}\right)$ is displacement operator, and $\rho_{\mathrm{th}}\left[(1-\eta_{d}+\nu_{\mathrm{el}})/{\eta_{d}}\right]$ is thermal state with mean photon number $(1-\eta_{d}+\nu_{\mathrm{el}})/{\eta_{d}}$. 

The secret key rate optimization based on the method in\ \cite{lin2020trusted} as shown in Appendix\ \ref{C}. Being different with\ \cite{lin2020trusted}, state of register $A$ with trusted source noise model is
\begin{equation}\label{rhoA}
	\begin{aligned} 
		&\rho_{A}=\mathrm{Tr}_B(\rho_{AB})=\frac{1}{\pi\bar{n}_s}\sum_{\scriptstyle i,j=0}^{3}\sqrt{p_{i}p_{j}}\;\;|i\rangle\langle j|_{A}\int_{\mathbb{C}}\exp\left(-\frac{|\beta|^{2}}{\bar{n}_s}\right)\\
		&\langle\sqrt{\eta}\alpha_{j}+\sqrt{1-\eta}\beta|\sqrt{\eta}\alpha_{i}+\sqrt{1-\eta}\beta\rangle d^{2}\beta,
	\end{aligned}
\end{equation}
which forms a different constraint for the key rate optimization problem in Eq.\ (\ref{opt}).

In experiments, the parameters of experimental equipment, such as the relative intensity noise of laser, extinction ratio of modulator, signal voltage and specific deviation of DAC can be used to calculate the source excess noise by Eqs. (\ref{eq:0})-(\ref{eq:3}). The trusted source noise can be separate with excess noise, and then our theoretical model can be used to calculate secret key rate.

\section{\label{set4}Simulations}
In this section, we initially present the channel model within the simulation framework and compute the simulated statistics. Subsequently, we simulate the QPSK modulated CV-QKD protocol, comparing three types of sources noise, i.e., ideal noiseless, untrusted, and trusted source noise, respectively.
\subsection{Simulation method}
To demonstrate the protocol's performance, we simulate the quantum channel as a phase-invariant Gaussian channel with transmissivity $\eta_t$ and excess noise $\xi$, where $\eta_{t}=10^{-\alpha L/10}$ for transmission distance $L$ in kilometers, $\alpha=0.2\ \text{dB/km}$ is the attenuation coefficient, and the excess noise is defined as
\begin{equation}\label{eq:0000}
\xi={\frac{(\Delta q_{\mathrm{obs}})^{2}}{(\Delta q_{\mathrm{vac}})^{2}}}-1
\end{equation}
where $(\Delta q_{\mathrm{vac}})^2=N_0=1/2$ is the variance of shot noise for $q$ quadrature, and $(\Delta q_{\mathrm{obs}})^2$ is $q$ quadrature's variance of the signal state measured by Alice. 

In this situation, simulated statistics and estimation of error-correction cost can be calculated. The simulated state $\sigma_{B}^{x}$ conditioned on selecting $x$ is a displaced thermal state, and its Wigner function is
\begin{equation}\label{eq:0000}
W_{\sigma_{B}^{x}}(\gamma)=\frac{1}{\pi}\frac{1}{\frac{1}{2}(1+\eta_{t}\eta_{s}\xi)}\exp\left[-\frac{|\gamma-\sqrt{\eta_{t}\eta_{s}}\alpha_{x}^{\prime}|^{2}}{\frac{1}{2}(1+\eta_{t}\eta_{s}\xi)}\right],
\end{equation}
where
\begin{equation}\label{eq:0000}
\alpha_{x}^{\prime}=\frac{1}{\pi\overline{{{n}}}_s}\int_{\mathbb{C}}\exp{(-|\beta|^{2}/\overline{{{n}}}_s)(\sqrt{\eta}\alpha_{x}+\sqrt{1-\eta}\beta)d^{2}\beta},
\end{equation}
which can be quickly calculated numerically.
Bob employs heterodyne measurement with trusted detector noise which is expressed by POVM ${G_y}$ and probability density function $P(y|x)$ of measurement result $y$, conditioned on Alice's selection $x$,
\begin{equation}\label{eq:0000}
P(y|x)=\frac{1}{\pi\left(1+\frac12\eta_{d}\eta_{t}\eta_{s}\xi+\nu_{\mathrm{el}}\right)}\exp\left[-\frac{|y-\sqrt{\eta_{d}\eta_{t}\eta_{s}}\alpha_{x}^{\prime}|^2}{1+\frac12\eta_{d}\eta_{t}\eta_{s}\xi+\nu_{\mathrm{el}}}\right].
\end{equation}

Expectation values of the observables in Eq.(\ref{obs}) can be written as
\begin{equation}\label{eq:0000}
\begin{cases}
\begin{aligned} 
\hspace{0.5em}
&\langle\hat{F}_{Q}\rangle_{x}=\sqrt{2\eta_{d}\eta_{t}\eta_{s}}\mathrm{Re}(\alpha_{x}^{\prime}),\\
&\langle\hat{F}_{P}\rangle_{x}=\sqrt{2\eta_{d}\eta_{t}\eta_{s}}\mathrm{Im}(\alpha_{x}^{\prime}),\\
&\langle\hat{S}_{Q}\rangle_{x}=2\eta_{d}\eta_{t}\eta_{s}\mathrm{Re}(\alpha_{x}^{\prime})^{2}+1+\frac{1}{2}\eta_{d}\eta_{t}\eta_{s}\xi+\nu_{\mathrm{el}},\\
&\langle\hat{S}_{P}\rangle_{x}=2\eta_{d}\eta_{t}\eta_{s}\mathrm{Im}(\alpha_{x}^{\prime})^{2}+1+\frac{1}{2}\eta_{d}\eta_{t}\eta_{s}\xi+\nu_{\mathrm{el}},\\
\end{aligned}
\end{cases}
\end{equation}
which are used as input of the SDP problem. In order to solve the SDP problem, we use the numerical framework in\ \cite{coles2016numerical,winick2018reliable}.

\begin{figure}[t]
	\centering
	\includegraphics[width=9cm]{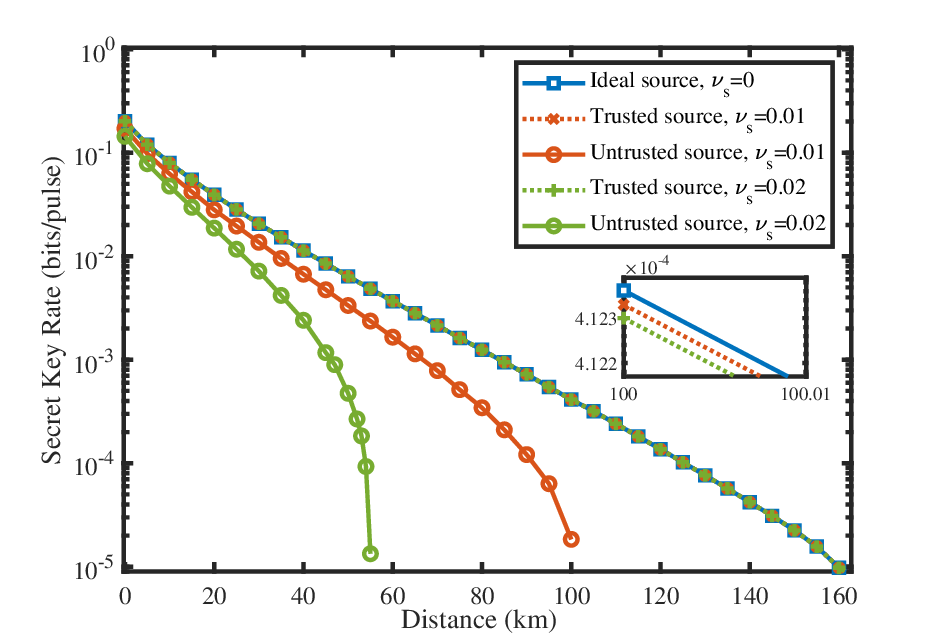}% Here is how to import EPS art
	\caption{\label{trust.vs.untrust}
		Comparisons of secret key rate versus distance of QPSK modulated CV-QKD with source noise $\nu_{s}=0.01$ and $\nu_{s}=0.02$ against ideal source $(\nu_{s}=0)$ under both trusted and untrusted models. Excess noise $\xi=0.02$, reconciliation efficiency $\beta = 0.956$, coherent-state amplitude $\alpha=0.6$, and postselection parameter $\Delta_a=0$.
	}
\end{figure}

\begin{figure}[t]
	\centering
	\includegraphics[width=9cm]{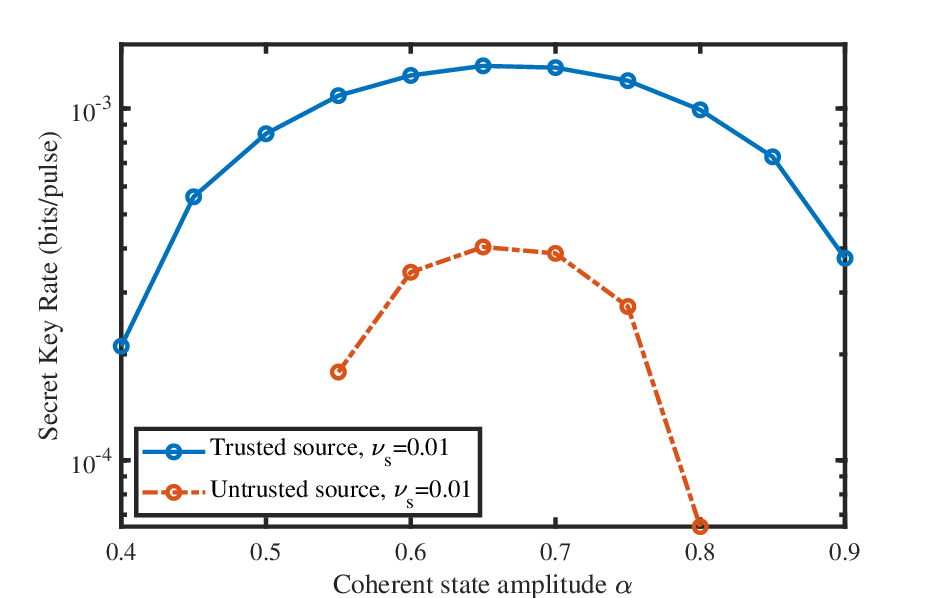}% Here is how to import EPS art
	\caption{\label{alpha_80km}
		Comparison of secret key rate versus coherent state amplitude $\alpha$ of QPSK modulated CV-QKD with imperfect source noise $\nu_{s}=0.01$, trusted and untrusted, respectively. Excess noise $\xi=0.02$, reconciliation efficiency $\beta = 0.956$, transmission distance $L=80$ km, and postselection parameter $\Delta_a=0$.
	}
\end{figure}

\subsection{Simulation results}
In this subsection, we first simulate the discrete-modulated CV-QKD with imperfect source in the case of ideal detector $(\nu_{el}=0,\eta_{d}=1)$. After that, we introduce imperfect detector with trusted noise $(\nu_{el}>0, \eta_{d}<1)$, and show the secret key rates. Furthermore, we optimize the coherent state amplitude, the optimal amplitude can serve as a valuable guideline for practical experimental implementations. Finally, effect of postselection on secret key rate is studied.

\subsubsection{Comparing source noise models}
In order to evaluate source noise's effect, we compare the performance of QPSK modulated CV-QKD protocol with ideal noiseless  source, trusted source noise, and untrusted source noise, respectively, as illustrated in Fig.\ \ref{trust.vs.untrust}, \ref{alpha_80km}.

\begin{figure}[t]
	\centering
	\includegraphics[width=9cm]{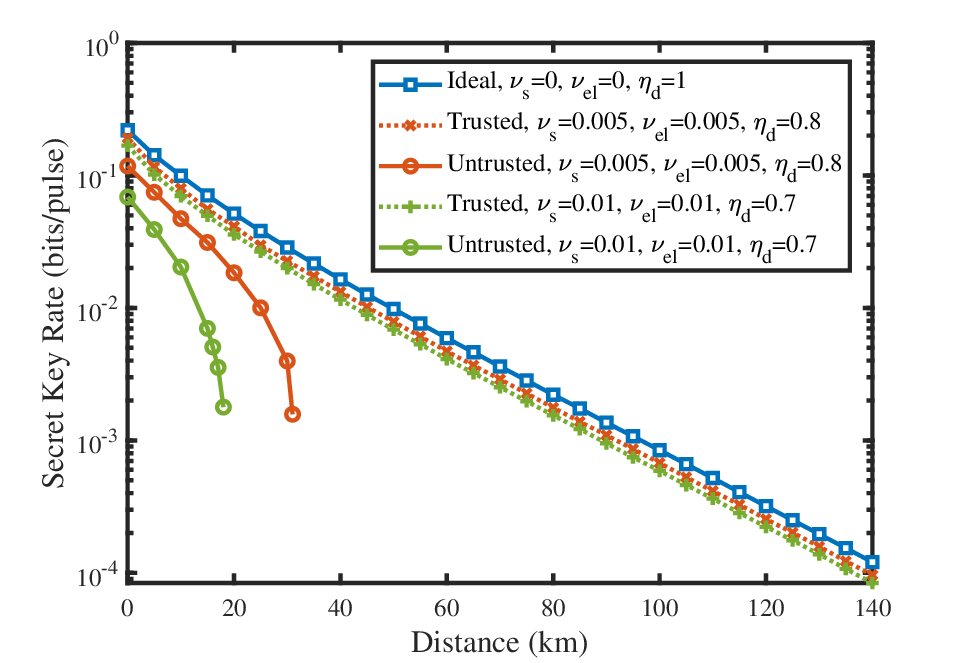}% Here is how to import EPS art
	\caption{\label{trust0}
		Comparisons of secret key rate versus distance of QPSK modulated CV-QKD with ideal, trusted and untrusted noise, respectively. Trusted detector noise is included. Excess noise $\xi=0.01$, reconciliation efficiency $\beta = 0.956$, coherent-state amplitude $\alpha=0.6$, and postselection parameter $\Delta_a=0$.
	}
\end{figure}

\begin{figure}[t]
	\centering
	\includegraphics[width=9cm]{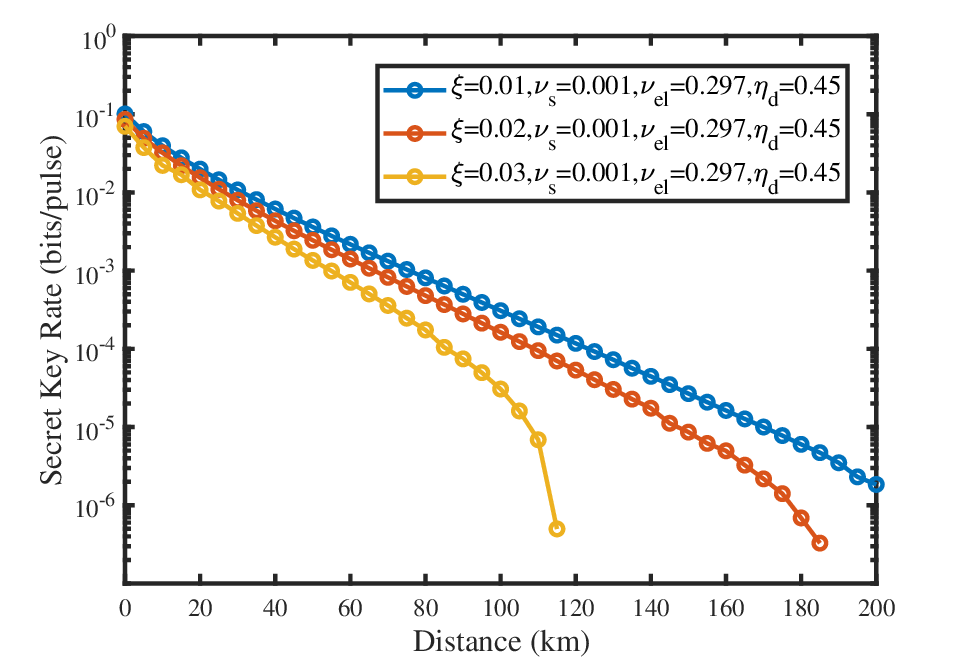}% Here is how to import EPS art
	\caption{\label{trust2}
		Comparisons of secret key rate versus distance of QPSK modulated CV-QKD with different excess noise. Source noise $\nu_{s}=0.001$, detector noise $\nu_{el}=0.297$, detection efficiency $\eta_{d}=0.45$,  reconciliation efficiency $\beta = 0.956$, coherent-state amplitude $\alpha=0.65$ and postselection parameter is $\Delta_a=0$. 
	}
\end{figure}

In Fig.\ \ref{trust.vs.untrust}, secret key rate omitting detector noise against transmit distance is compared. We take excess noise $\xi$ to be 0.02, error-correction efficiency $\beta$ = 0.956\ \cite{ma2023practical}, without considering the postselection progress. We set the source noise value $\nu_{s}=0.01$ and $0.02$, respectively. Simulation results indicate that, even with source noise of 0.02, performance of the protocol with trusted source noise is almost equivalent to that with ideal source. In contrast, when source noise cannot be trusted, maximum transmission distance of the protocol is less than 60 km under source noise $\nu_{s}=0.02$. However, when it can be trusted, it can be transmitted to more than 160 km. The result shows that the proposed trusted noise model almost eliminates impact of imperfect source on performance of discrete-modulated CV-QKD. Compared with untrusted source, the performance is greatly improved, both on transmission distance and secret key rate. Moreover, the performance improvement becomes more significant as the source noise increases.

In Fig.\ \ref{alpha_80km}, the coherent state amplitude is optimized for the QPSK modulated CV-QKD protocol in 80 km with trusted source and untrusted source noise, respectively, where source noise $\nu_{s}=0.01$ and other parameters are the same as in Fig.\ \ref{trust.vs.untrust}. Simulation results show that in this condition, the optimal coherent state modulation amplitude is about 0.65, no matter the source noise is trusted or not. However, when the source noise can be trusted, secret key rate of the protocol is more than $10^{-3}$ bits per pulse nearly an order of magnitude larger than its untrusted source noise counterpart.

\begin{figure}[t]
	\centering
	\includegraphics[width=9cm]{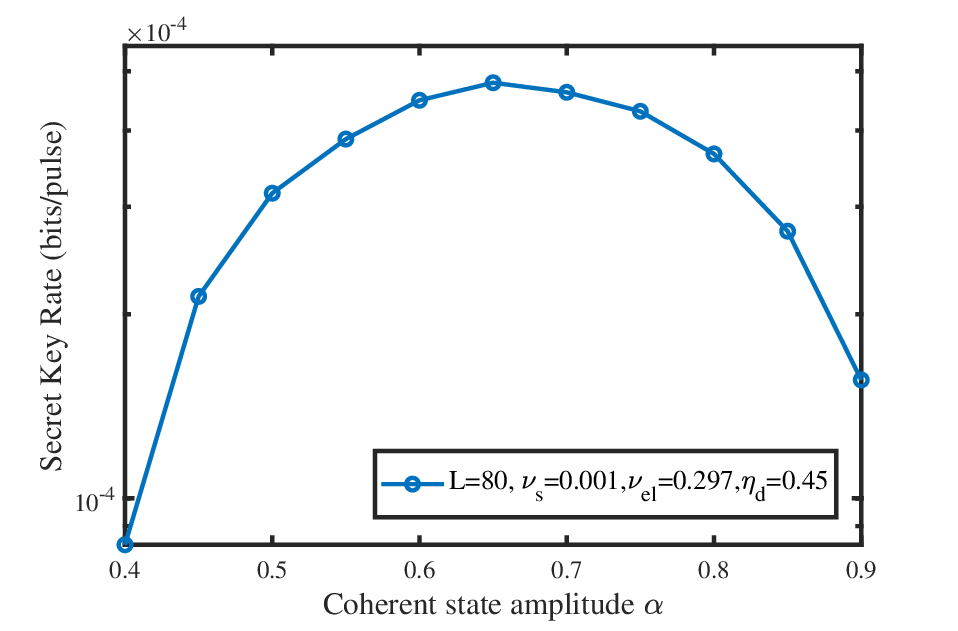}% Here is how to import EPS art
	\caption{\label{alpha2}
		Optimizing coherent state amplitude $\alpha$ of QPSK modulated CV-QKD with excess noise $\xi=0.02$, source noise $\nu_{s}=0.001$, detector noise $\nu_{el}=0.297$, detection efficiency $\eta_{d}=0.45$, reconciliation efficiency $\beta = 0.956$, and postselection parameter $\Delta_a=0$. When the source and detector noise is untrusted, secret key rate is 0.
	}
\end{figure}

\subsubsection{Including detector noise}
Next, we consider actual situation in the experiments, further imperfect detector, and use real parameters to simulate the QPSK modulated CV-QKD protocol with trusted detector noise model\ \cite{lin2020trusted}.

In Fig.\ \ref{trust0}, we compare secret key rate of the protocol with ideal, trusted and untrusted noise, respectively (Trusted noise means that both the source noise and detector noise are trusted, and untrusted noise means that both noise are untrusted). For noise amplitude, two situations with $\nu_s=0.005$, $\nu_{el}=0.005$, $\eta_{d}=0.8$, and $\nu_s=0.01$, $\nu_{el}=0.01$, $\eta_{d}=0.7$ are considered to represent high and low noise status. Compared with ideal sources and detectors, key rate under trusted noise is slightly lower, mainly due to imperfect detection efficiency. However, when the noise from source and detector become untrusted, the protocol performance rapidly declines, primarily since the noise is factored into the excess noise, and the equivalent excess noise on the source side $\xi=\nu_s+\nu_{el}/\eta_t$. When transmission distance increases, excess noise significantly rises, leading to a rapid decrease of secret key rate. The trusted models effectively addresses this issue, and enable the protocols to maintain high performance.

In Fig.\ \ref{trust2}, we show secret key rate versus distance of with different excess noise, where source noise $\nu_s=0.001$, detector noise $\nu_{el}=0.297$, and detection efficiency $\eta_{d}=0.45$\ \cite{wang2022sub}. When they cannot be trusted, the protocol cannot generate secret keys. When source noise and detector noise can be trusted, the simulation outcomes indicate that, under the given parameter, the protocol can transmit over 200 km when excess noise is 0.01. However, as excess noise increases to 0.02, secret key rate experiences a notable decline once transmission distance surpasses 180 km. Furthermore, with excess noise of 0.03, the protocol's transmission limit is reduced to approximately 120 km.

\begin{figure}[t]
	\centering
	\includegraphics[width=9cm]{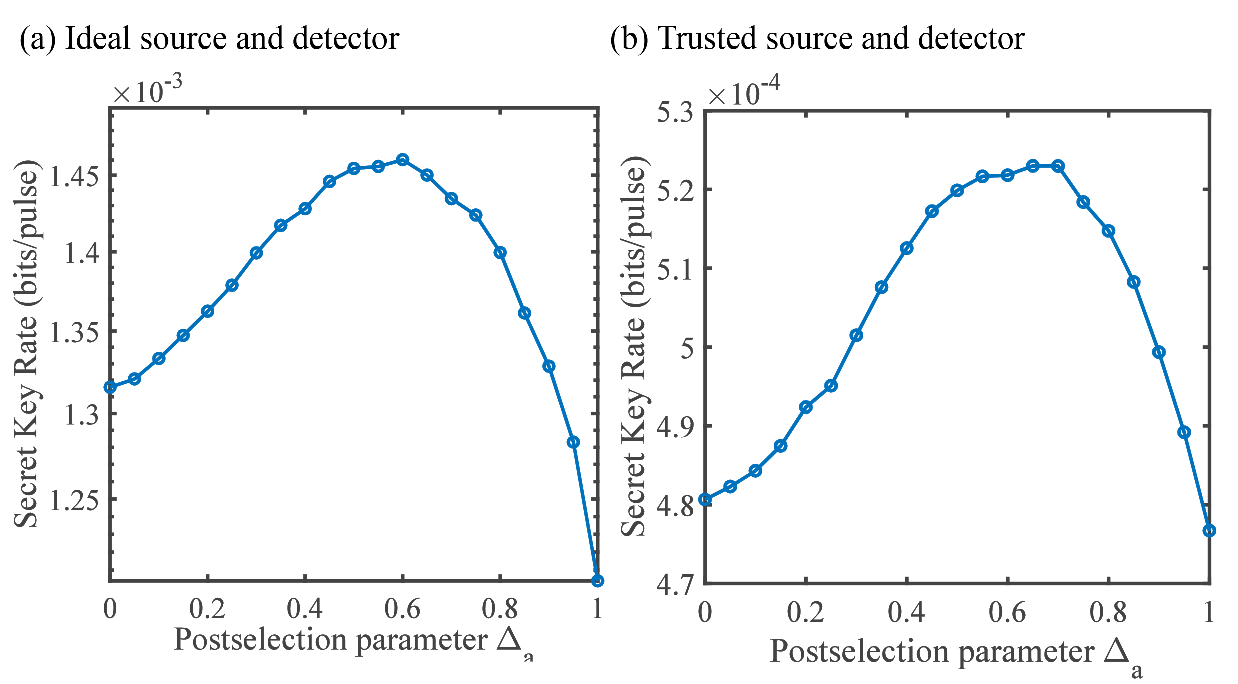}% Here is how to import EPS art
	\caption{\label{delta}
		Optimizing postselection parameter $\Delta_a$ of QPSK modulated CV-QKD. (a) Ideal source and detector. (b) Trusted source and detector. Both with distance $L=80$ km, excess noise $\xi=0.02$, source noise $\nu_{s}=0.001$, detector noise $\nu_{el}=0.297$, detection efficiency $\eta_{d}=0.45$, and reconciliation efficiency $\beta = 0.956$. 
	}
\end{figure}

In Fig.\ \ref{alpha2}, the coherent state amplitude is optimized for QPSK modulated CV-QKD protocol in 80 km with trusted source noise and detector noise, excess noise $\xi=0.02$, and other parameters are the same as Fig.\ \ref{trust2}. Simulation results show that the optimal coherent state amplitude is about 0.65 in this condition which is almost equal to the situation without trusted noise model in Fig.\ \ref{alpha_80km}. The secret key rate with optimal coherent state amplitude is about $4\times10^{-4}$ bits per pulse.

\begin{figure}[t]
	\centering
	\includegraphics[width=8cm]{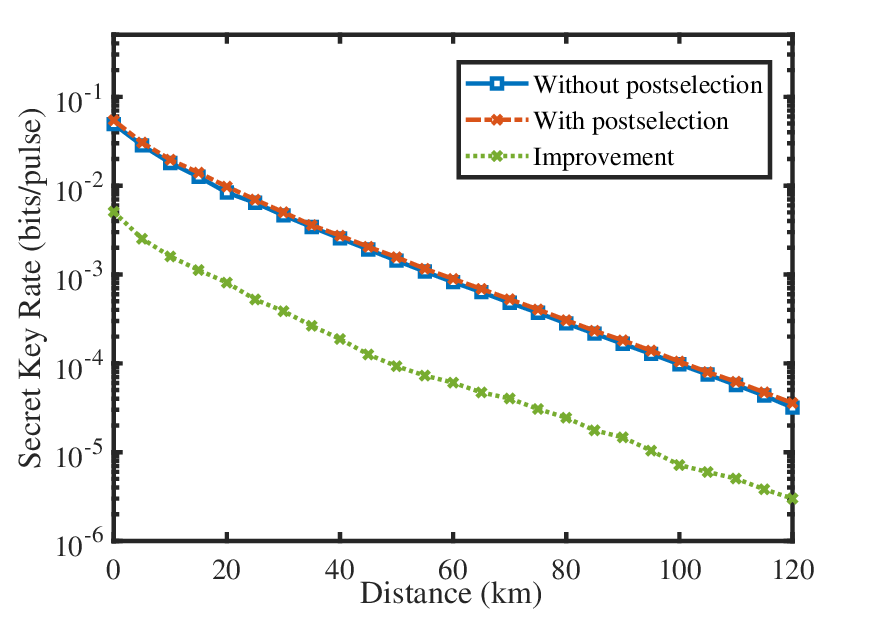}% Here is how to import EPS art
	\caption{\label{different}
		Comparison of secret key rate versus distance of QPSK modulated CV-QKD with and without postselection, source noise $\nu_{s}=0.001$, detector noise $\nu_{el}=0.297$, detection efficiency $\eta_{d}=0.45$ and reconciliation efficiency $\beta = 0.956$, coherent-state amplitude $\alpha=0.65$, and the postselection parameter is optimized.
	}
\end{figure}

\subsubsection{Including postselection}
Then we consider the influence of trusted source and detector noise models on postselection parameter, and demonstrate the enhancement of secret key rate in QPSK modulated CV-QKD with trusted source and detector through postselection.

In Fig.\ \ref{delta}, postselection parameter of QPSK modulated CV-QKD in 80 km with trusted source and detector noise is optimized, in parallel with the ideal noiseless case, where coherent state amplitude is $\alpha=0.65$ and other parameters are the same as Fig.\ \ref{alpha2}. Simulation results show that, for ideal case as shown in Fig.\ \ref{delta}(a), the optimal postselection parameter is about 0.65, and secret key rate can increase by up to approximately $7 \%$ compared to the rate without postselection($\Delta_{a}=0$). In Fig.\ \ref{delta}(b), with trusted noise model, the optimal postselection parameter is about 0.7, and secret key rate can be improved from $4.8\times10^{-4}$ ($\Delta_{a}=0$) to $5.2\times10^{-4}$ bits per pulse, which is about $8 \%$ higher. Overall, trusted noise model would not change much the optimal postselection parameter, neither performance improvement of postselection.

In Fig.\ \ref{different}, we demonstrate the performance improvement of optimal postselection at different distances, under noise parameters the same as Fig.\ \ref{trust2}. Simulation results show that postselection can improve the secret key rate by about $7 \%$ at various distances. 

In summary, simulation results exhibit significant enhanced performance of the QPSK modulated CV-QKD protocol when source and detector noise can be trusted, compared with the untrusted counterpart. In addition, applying trusted noise models have little impact on post selection, and reasonable postselection can further improve the performance of discrete modulated CV-QKD system.
%All in all, trusted source and detector noise models play an important role in reducing the impact of source and detector noise on system performance in practical experiments.

\section{\label{set5}Discussion and Conclusion}
Over the past five years, theoretical security of discrete-modulated CV-QKD has been progressively refined. The modeling of trusted noise in practical system is an important part of the practical security of CV-QKD, which can avoid the loss of secret key rate caused by mistakenly listing trusted noise as untrusted. Trusted noise model makes sense since the devices are inside the system and cannot be manipulated. It is worth noting that there may be some quantum hacking attacks in the future which can break this assumption. For example, the modulation noise of IQ modulator can be trusted in many cases, but if Eve could control the driver electrical signal of IQ modulator, this assumption is bad. Once such attacks are proposed, these parts of noise cannot be regarded as trusted noise. If communication parties still consider them as trusted noise, the overestimation of secret key rate will have a serious impact on system security. In this case, these parts of noise need to be regarded as untrust, and our trusted modeling is still applicable to other source noise. Meanwhile, some countermeasures are also feasible, such as electromagnetic shielding, to keep the source noise trusted.

In this contribution, we examine the challenges posed by imperfect source in discrete-modulated CV-QKD system and proposed a trusted source noise model. Our work can readily be applied to the finite-size regime which ensure composable security, and protocols using higher-order modulation.   The results show that, compared with previous treatment, which classifies all source noise into the excess noise controllable by eavesdroppers, this model more accurately evaluate the ability of eavesdroppers and improve the secret key rate of discrete-modulated CV-QKD systems, which takes an important step towards practical application of discrete-modulated CV-QKD.

\begin{acknowledgments}
This research was supported by the National Natural Science Foundation of China (62001044), the Basic Research Program of China (JCKY2021210B059), the Equipment Advance Research Field Foundation (315067206), the Fund of State Key Laboratory of Information Photonics and Optical Communications, and the Fundamental Research Funds for Central Universities, China (2023RC28).
\end{acknowledgments}

\appendix

\section{\label{A}Protocol description of QPSK modulated CV-QKD with imperfect source}
The prepare-and-measure version of practical QPSK modulated protocol as follow:

$(1)$ State preparation. For each round, Alice employs a laser, which is then integrated into the modulator, facilitating the emission of a coherent state ${\vert{\varphi_{k}}\rangle}$ chosen from the set $\left\{\vert\alpha\rangle,\vert-\alpha\rangle,\vert i\alpha\rangle,\vert-i\alpha\rangle\right\}$ according to probability $p_{k}=1/4$, where $\alpha\in\mathbb{R}$ is the amplitude of the coherent states. The modulator is driven by a DAC. It should be noted that these devices are imperfect and contain non ideal factors as shown in Sec. \ref{set2}. The coherent state is sent to Bob through the quantum channel which is controlled by Eve. 

$(2)$ Measurement. After receiving the state sent from Alice, Bob uses a heterodyne detector with trusted noise and obtains the measurement result $y\in\mathbb{C}$.

\begin{figure}[t]
	\centering
	\includegraphics[width=6cm]{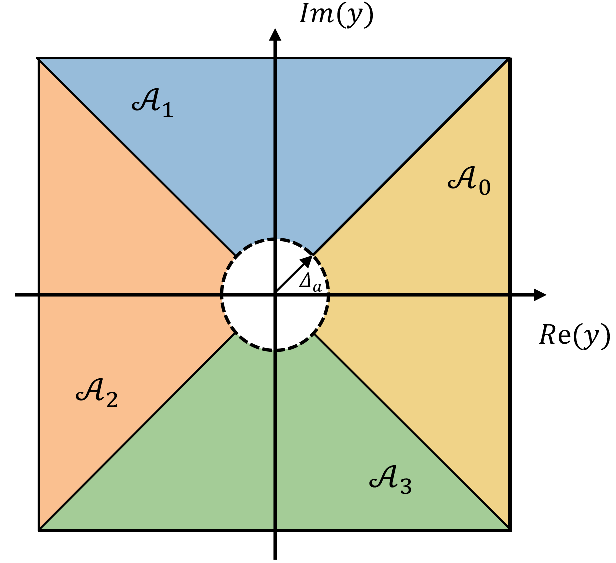}% Here is how to import EPS art
	\caption{\label{QPSK}
		Schematic diagram of Bob's key map step for the measurement results $y$. Each region ${\cal A}_{z}$ represents a key map value $j$. During the postselection process of data, the measurement result obtained from the central disk with a radius of $\Delta_a$ is disregarded and instead assigned the symbol $\perp$.
	}
\end{figure}

The physical process of the prepare-and-measure version of practical QPSK modulated protocol is shown in Fig.\ \ref{pm1} (a), and the excess noise introduced in Sec.\ \ref{set2} is considered. The excess noise makes the variance of quadrature increase from $N_0$ to $N_{0}(1+\xi)$, and its phase-space representation is shown in Fig.\ \ref{pm1} (b). After the physical process of the protocol, Alice and Bob perform post-processing through the classic channel, including announcement and sifting, parameter estimation, reverse reconciliation key map, error correction and privacy amplification. These processes are similar to those in the literature \cite{lin2019,lin2020trusted}. For the sake of integrity, we briefly introduce these steps:

$(3)$ Announcement and sifting. After N rounds of communication, Alice and Bob identify a small subset of test rounds $\tau_{\mathrm{test}}$ used for parameter estimation, use the remaining rounds $\tau_{\mathrm{key}}$ to generate keys. Following the sifting process, Alice obtains her string $\textbf{X}=(x_{1},...,x_{m})$ according to the following rule:

\begin{equation}
    \forall j\in[m], \ \ x_{j}=
    \begin{cases}
	        0 \ \ \text{if} \ \ |\psi_{f(j)}\rangle=|\alpha\rangle, \\
	        1 \ \ \text{if} \ \ |\psi_{f(j)}\rangle=|i\alpha\rangle, \\
	        2 \ \ \text{if} \ \ |\psi_{f(j)}\rangle=|-\alpha\rangle, \\
	        3 \ \ \text{if} \ \ |\psi_{f(j)}\rangle=|-i\alpha\rangle, \\
	    \end{cases}
\label{KD}
\end{equation}
where $m$ is the size of the set $\tau_{\mathrm{key}}$, $f$ is a function from $[m]$ to $\tau_{\mathrm{key}}$.

$(4)$ Parameter estimation. Alice and Bob carry out parameter estimation by revealing all information from the rounds designated by the test set $\tau_{\mathrm{test}}$. In order to conduct this analysis, they process the data by calculating the observable measurement, conditioned on each of the four states sent by Alice. These metrics enable them to place limitations on their joint state $\rho_{AB}$. Subsequently, they compute the secret key rate in accordance with the optimization problem. If their analysis indicates that secret keys cannot be produced, they terminate the protocol. Otherwise, they move forward.

$(5)$ Reverse reconciliation key map. Bob employs a key map progress to derive his raw key string. This map process transforms his measurement result $y_{k}$ into an element within a specific set $\left\{0, 1, 2 , 3, \perp\right\}$, Bob obtains his key string $\mathbf{Z}=(z_{1},...,z_{m})$ according to the rule as shown in Fig.\ \ref{QPSK},
\begin{equation}
    z_{j}=
    \begin{cases}
	        j \ \ \quad{\mathrm{if}}\,\theta\in{\bigg[}{\frac{(2j-1)\pi}{4}},{\frac{(2j+1)\pi}{4}}{\bigg)}\operatorname{and}|y|\geq\Delta_{a}, \\
	        \perp\quad{ \text{otherwise}}.
	
	    \end{cases}
\label{KD}
\end{equation}
where $\Delta_{a}$ is a postselection parameter and $j\in \left\{0,1,2,3\right\}$.

$(6)$ Error correction and privacy amplification. Alice and Bob use privacy amplification to diminish Eve's knowledge of their shared information by eliminating certain portion of their jointly held key.

\section{\label{B}Coupling of two coherent states through a beam splitter}
Considering that coherent states $|\alpha_{A'}\rangle$ and $|\alpha_{R}\rangle$ are coupled through a beam splitter with transmissivity $\eta$, the output coherent state is $|\alpha_{A''}\rangle$ and $|\alpha_{F}\rangle$. For the input coherent state $|\alpha_{A'}\rangle$, we can rewritten it with displace operator $\hat{D}(\alpha_{A''})$ as
\begin{equation}\label{eq:0000}
	|\alpha_{A^{\prime}}\rangle=\hat{D}(\alpha_{A^{\prime}})|0\rangle_{A^{\prime}},
\end{equation}
and the displace operator is
\begin{equation}\label{A2}
	\hat{D}(\alpha_{A'})=\mathrm{exp}\left(\alpha_{A'}\hat{\alpha}_{A'}^{\dagger}-\alpha_{A'}^{*}\hat{\alpha}_{A'}\right).
\end{equation}
The beam splitter operator can be expressed as
\begin{equation}\label{eq:0000}
	\hat{S}(\theta)=\exp[\theta(\hat{a}_{A^{\prime}}^{\dagger}\hat{a}_{R}-\hat{a}_{A^{\prime}}\hat{a}_{R}^{\dagger})]
\end{equation}
where parameter $\theta$ is related to transmittance $\eta_{s}$ by $\eta_{s}=1/(1+\rm{tan}^{2}\theta)$. It can also be written in matrix form as
\begin{equation}
	\begin{gathered}
		\begin{pmatrix} \hat{a}_{A''} \\ \hat{a}_{F} \end{pmatrix}
		= \begin{pmatrix} \sqrt{\eta} & \sqrt{1-\eta} \\ \sqrt{1-\eta} & -\sqrt{\eta}\end{pmatrix}
		\begin{pmatrix} \hat{a}_{A'} \\\hat{a}_{R} \end{pmatrix}
	\end{gathered}
\end{equation}
Thus we have
\begin{equation}\label{A5}
	\hat{a}_{A^{\prime}}=\sqrt{\eta}\hat{a}_{A''}+\sqrt{1-\eta}\hat{a}_{F},
\end{equation}
Substitute (\ref{A2}), (\ref{A5}) into
\begin{equation}\label{eq:0000}
	\begin{split} 
		\hat{S}(\eta)&\tilde{D}(\alpha_{A^{\prime}})\hat{S}^{\dagger}(\eta)\\
		=&\exp\left(\sqrt{\eta}\big(\alpha_{A^{\prime}}\hat{\alpha}_{A''}^{\dagger}-\alpha_{A^{\prime}}^{*}\hat{a}_{A''}\big)\right)\times\\
		&\exp\left(\sqrt{1-\eta}\big(\alpha_{A^{\prime}}\hat{a}_{F}^{\dagger}-\alpha_{A^{\prime}}^{*}\hat{a}_{F}\big)\right)
	\end{split}
\end{equation}
Let $\hat{A}=\alpha_{A^{\prime}}\hat{\alpha}_{A''}^{\dagger}-\alpha_{A^{\prime}}^{*}\hat{\alpha}_{A''}$ and $\hat{B}=\alpha_{A^{\prime}}\hat{a}_{F}^{\dagger}-\alpha_{A^{\prime}}^{*}\hat{a}_{F}$, the commutator between $\hat{A}$ and $\hat{B}$ can be calculated as
\begin{equation}\label{eq:0000}
	\left[\hat{A}, \hat{B}\right]=\left[\alpha_{A^{\prime}}\hat{\alpha}_{A''}^{\dagger}-\alpha_{A^{\prime}}^{*}\hat{\alpha}_{A''},\alpha_{A^{\prime}}\hat{\alpha}_{F}^{\dagger}-\alpha_{A^{\prime}}^{*}\hat{\alpha}_{F}\right]=0
\end{equation}
Using Baker‐Campbell‐Hausdorff formula
\begin{equation}\label{eq:0000}
	\begin{split} 
		&\hat{S}(\eta)\widehat{D}_{A^{\prime}}(\alpha_{A^{\prime}})\hat{S}(\eta)^{\dagger}\\
		&\hspace{2em}=\exp\!\left(\sqrt{\eta}\hat{A}\right)\exp\!\left(\sqrt{1-\eta}\hat{B}\right)\\
		&\hspace{2em}=\exp\left(\sqrt{\eta}(\alpha_{A^{\prime}}\hat{a}_{A''}^{\dagger}-\alpha_{A^{\prime}}^{*}\hat{a}_{A''})\right)\times\\
		&\hspace{3em}\exp\left(\sqrt{1-\eta}(\alpha_{A^{\prime}}\hat{a}_{F}^{\dagger}-\alpha_{A^{\prime}}^{*}\hat{a}_{F})\right),\\
		&\hspace{2em}=\hat{D}_{A''}\big(\sqrt{\eta}\alpha_{A^{\prime}}\big)\hat{D}_{F}\big(\sqrt{1-\eta}\alpha_{A^{\prime}}\big)
	\end{split}
\end{equation}
Therefore after passing the beam splitter,
\begin{equation}\label{eq:0000}
	\begin{split} 
		|\alpha_{A^{\prime}}\rangle&\rightarrow\tilde{D}_{A''}(\sqrt{\eta}\alpha_{A^{\prime}})\tilde{D}_{F}(\sqrt{1-\eta}\alpha_{A^{\prime}})|0\rangle_{A^{\prime}}\\
		&=\left|\sqrt{\eta}\alpha_{A^{\prime}}\right>_{A''}\left|\sqrt{1-\eta}\alpha_{A^{\prime}}\right>_{F}.
	\end{split}
\end{equation}
Similarly,
\begin{equation}\label{eq:0000}
	\left|\alpha_{R}\right\rangle\rightarrow\left|\sqrt{1-\eta}\alpha_{R}\right\rangle_{A''}\left|-\sqrt{\eta}\alpha_{R}\right\rangle_{F}.
\end{equation}
Thus, we can obtain that the output quantum state is
\begin{equation}\label{eq:0000}
	\vert\alpha_{A''}\rangle=\vert\sqrt{\eta}\alpha_{A^{\prime}}+\sqrt{1-\eta}\alpha_{R}\rangle_{A''}.
\end{equation}

\section{\label{C}Optimization of secret key rate}
The secret key rate optimization method is based on\ \cite{lin2020trusted}. Given annihilation operator $\hat{a}$  and creation operator $\hat{a}^{\dagger}$ satisfying the basic commutation relation $[{\hat{a}},{\hat{a}}^{\dagger}]=1$, the quadrature operators $\hat{q}$ and $\hat{p}$ is defined as
\begin{equation}\label{eq:0000}
\hat{q}=\frac{1}{\sqrt{2}}(\hat{a}^{\dagger}+\hat{a}),\quad\hat{p}=\frac{i}{\sqrt{2}}(\hat{a}^{\dagger}-\hat{a}).
\end{equation}
%In this case, the variance of shot noise $N_0$ can be calculated as $\frac{1}{2}$, and the commutation relation between the quadrature operators is $[{\hat{q}},{\hat{p}}]=i$.

Next, we discuss how to calculate the secret key rate of the protocol with the imperfect source model based on the entanglement-based scheme. The Devetak-Winter formula can be rewritten in the form\ \cite{coles2016numerical} 
\begin{equation}\label{eq:0000}
R^{\infty}=\operatorname*{min}_{\rho_{A B}\in{\bf S}}D\Bigl({\cal G}(\rho_{A B})\vert\vert {\cal Z}\lbrack{\cal G}(\rho_{A B})\rbrack\Bigr)-p_{\mathrm{pass}}\delta_{\mathrm{EC}},
\end{equation}
where ${\cal G}$ is a CPTP map which outlines several classical post-processing procedures of the protocol, ${\cal Z}$ is a pinching quantum channel which is used to access results of the key map. $D(\rho||\sigma)=\operatorname{Tr}(\rho\log_{2}\rho)-\operatorname{Tr}(\rho\log_{2}\sigma)$ is the quantum relative entropy between the quantum states $\rho$ and $\sigma$. ${\bf S}$ is the set of density matrices satisfying experimental constraints. 
The key of the problem lies in the first term of the formula, namely optimization problem $\operatorname*{min}_{\rho_{A B}\in{\bf S}}D\Bigl({\cal G}(\rho_{A B})\vert\vert {\cal Z}\lbrack{\cal G}(\rho_{A B})\rbrack\Bigr)$. The optimization problem can be expressed as
\begin{equation}\label{opt}
\begin{aligned} 
&\text{minimize} \operatorname*{min}_{\rho_{A B}\in{\bf S}}D\Bigl({\cal G}(\rho_{A B})\vert\vert {\cal Z}\lbrack{\cal G}(\rho_{A B})\rbrack\Bigr)\\
&\text{subject to} \\
&\hspace{2em}
\begin{cases}
\vspace{2pt}
\hspace{0.5em}\mathrm{Tr}[\rho_{AB}(|x\rangle\langle x|_{A}\otimes{\hat{F}}_{Q})]=p_{x}\langle{\hat{F}}_{Q}\rangle_{x},\\
\vspace{2pt}
\hspace{0.5em}\mathrm{Tr}[\rho_{AB}(|x\rangle\langle x|_{A}\otimes{\hat{F}}_{P})]=p_{x}\langle{\hat{F}}_{P}\rangle_{x},\\
\vspace{2pt}
\hspace{0.5em}\mathrm{Tr}[\rho_{AB}(|x\rangle\langle x|_{A}\otimes{\hat{S}}_{Q})]=p_{x}\langle{\hat{S}}_{Q}\rangle_{x},\\
\vspace{2pt}
\hspace{0.5em}\mathrm{Tr}[\rho_{AB}(|x\rangle\langle x|_{A}\otimes{\hat{S}}_{P})]=p_{x}\langle{\hat{S}}_{P}\rangle_{x},\\
\vspace{2pt}
\hspace{0.5em}\mathrm{Tr}[\rho_{AB}]=1,\\
\vspace{2pt}
\hspace{0.5em}\mathrm{Tr}_{B}[\rho_{A B}]=\rho_{A}
\end{cases}
\end{aligned}
\end{equation}
where index $x\in\left\{0,1,2,3\right\}$, $\langle{\hat{F}}_{Q}\rangle, \langle{\hat{F}}_{P}\rangle, \langle{\hat{S}}_{Q}\rangle, \langle{\hat{S}}_{P}\rangle_{x}$ is expectation values of operators $\hat{F}_{Q}, \hat{F}_{P}, \hat{S}_{Q}, \hat{S}_{P}$ for the conditional state $\rho_{B}^{x}$ respectively. The observable operators are
\begin{equation}\label{obs}
\begin{cases}
\begin{aligned} 
\hspace{0.5em}
&\hat{F}_{Q}=\int\frac{y+y^{*}}{\sqrt{2}}G_{y}d^{2}y,\\
&\hat{F}_{P}=\int\frac{i(y^{*}-y)}{\sqrt{2}}G_{y}d^{2}y,\\
&\hat{S}_{Q}=\int\left(\frac{y+y^{*}}{\sqrt{2}}\right)^{2}G_{y}d^{2}y,\\
&\hat{S}_{P}=\int\left[\frac{i(y^{*}-y)}{\sqrt{2}}\right]^{2}G_{y}d^{2}y.
\end{aligned}
\end{cases}
\vspace{2pt}
\end{equation}
In addition, in the last constraint, $\rho_{A}$ is calculated as Eq.\ (\ref{rhoA}).
In order to perform postselection, the region operators is defined as
\begin{equation}\label{eq:0000}
R_{z}=\int_{y\in {\cal{A}}_{z}}G_{y}d^{2}y,
\end{equation}
where ${\cal{A}}_{z}$ is the region of integration corresponds to the regions shown in Fig.\ \ref{QPSK}. For reverse reconciliation, the CPTP map ${\cal G}(\sigma)=K\sigma K^{\dagger}$ with input state $\sigma$, where $K$ is the Kraus map as
\begin{equation}\label{eq:0000}
K=\sum_{z=0}^{3}|z\rangle_{R}\otimes\mathrm{I}_{A}\otimes(\sqrt{R_{z}})_{B}.
\end{equation}
The pinching quantum channel is described by
\begin{equation}\label{eq:0000}
{\mathcal{Z}}(\sigma)=\sum_{j=0}^{3}(|j\rangle\langle j|_{R}\otimes\mathrm{I}_{A B})\sigma(|j\rangle\langle j|_{R}\otimes\mathrm{I}_{A B}).
\end{equation}

For the second term of the secret key rate formula, $p_{\rm pass}$ is the sifting probability, and $\delta_{\mathrm{EC}}$ is the cost of error correction. This part is classical and computable. For the reverse reconciliation scheme, The cost of error correction $\delta_{\mathrm{EC}}$ can be described as
\begin{equation}\label{eq:0000}
	\delta_{\mathrm{EC}}=H(\mathbf{Z})-\beta I(\mathbf{X};\mathbf{Z}),
\end{equation}
where $H(\mathbf{\textbf{Z}})$ represents the classical information entropy entropy of the raw key $\textbf{Z}$, $\beta$ signifies the reconciliation efficiency, and $I(\textbf{X}; \textbf{Z})$ denotes the classical mutual information.
To sum up, the optimization problem can be solved, subsequently enabling derivation of secret key rate.

\bibliography{ref0}
\end{document}